\documentstyle[epsfig,12pt]{article}
\begin{document}

\title{ A Multi-Time Scale Non-Gaussian Model of Stock Returns}

\author{Lisa Borland \\Evnine-Vaughan Associates, Inc.\\ 456 Montgomery Street, Suite 800\\
San Francisco, CA 94104, USA\\ lisa@evafunds.com\\}

\maketitle

\begin{abstract}
We propose a  stochastic process for stock movements that, with just one source of Brownian noise,
 has an instantaneous volatility that rises from a type of  statistical 
feedback across  many time scales.  This results in a stationary non-Gaussian process which  captures
many features  observed in time series of real stock returns. These include
volatility clustering, a  kurtosis which decreases slowly over time together with a  close to log-normal distribution of 
instantaneous volatility. We calculate the rate of decay of volatility-volatility correlations, which depends on the 
strength of the memory in the system and fits well to  empirical observations.
\end{abstract}

\section{Introduction}

In the past years there has been a body of work investigating the statistical properties of 
financial data \cite{bouchaudpotters,stanley}, in particular of time series of stock returns. Such data is available on the smallest of time
scales, namely tick by tick, ranging to years and years. Certain properties appear to be extremely stable across
different stocks and different time horizons. For example, histograms of stock returns exhibit fat tails
such that the cumulative distribution decays as a power law with exponent $-3$. This behaviour is extremely stable,
decaying slowly towards a Gaussian distribution as the time scale of the returns is increased. 
Time series of returns exhibit bursts of correlated volatility, in the sense that there will be regions of higher
or lower volatility, together with a persistence in the sense that the volatility exhibits a rather significant correlation
over time. In addition, the distribution of empirical volatilities can be fit very well by a close-to log-normal distribution.
 
The standard log-normal model of stock returns, inspired by earlier work due to Bachelier \cite{Bachelier} and  used for example by Black, Scholes and Merton \cite{Black&Scholes,Merton} for the purpose of 
option pricing,  has had great success as the basic building block for modeling stock returns. However, none of the
interesting features of financial time series which we mentioned above are captured by such a model.
Instead, a variety of other models have been proposed in order to  more realistically describe financial data 
\cite{Heston,bacry,muzy,Pochart,bouchaudleverage,peinke,gunaratne}.
For example, stochastic volatility models \cite{Heston,Fouque,CGMY,Perello}  capture well the volatility clustering and fat tails, except that these tails decay way too quickly towards a Gaussian distribution.
Multifractal models have also enjoyed great success in describing many of the stylized facts \cite{bacry,muzy,Pochart}, although they contain an additional
 source of noise,  in contrast to the model we present in this paper. It is thus still an exciting open question, to probe the dynamics giving rise 
to price formation both on a fundamental level, namely by studying the dynamics of the order books
\cite{jpprice,doyne,challet} as well as on a more phenomenological level, namely by understanding the possible types of stochastic
dynamics that could underlie price formation. The current paper will address this issue along the latter lines.

We recently proposed modeling the driving noise of stocks by a statistical
feedback process of the type presented in \cite{pre_borland}.
Based on that model, we were able to quite well capture many features of stock option markets,
lending  weight to the model \cite{prl_borland,qf_borland,borlandbouchaud}. However, as pointed out in those papers,
the model is not an entirely realistic  one. The main reason is
that there is one single characteristic time  in that model, and 
in particular the effective volatility at each time is related to the conditional
probability of observing an outcome of the process at time t {\it given 
what was observed at time t = 0}. This is a shortcoming
of that model for one of real stock returns; in real markets, traders
drive the price of the stock based on their own trading horizon. 
But there are traders who react to each tick the stock makes, ranging to those
reacting to what they believe is relevant on the horizon of a  year or more,
and of course, there is the entire spectrum in-between. Therefore,
an optimal model of real price movements should attempt to capture this
existence of multi-time scales.

\section{ The Model of Stock Returns}

\subsection{The statistical feedback model}

First we summarize the original model, in a slightly different notation. If $S$ is the stock price, then 
$y = \ln S$ is the log-stock price. We proposed (\cite{qf_borland}) that
\begin{equation}
\label{eq:y}
d y = \mu dt + \sigma d \Omega
\end{equation}
with
\begin{equation}
d \Omega = P(\Omega\mid \Omega_0)^{\frac{1-q}{2}} d \omega
\end{equation}
Here, $\mu $ is the instantaneous log return, $\sigma$ is the volatility parameter, and $\omega$ is a standard Brownian noise such that
$\langle   d \omega(t) d\omega(t') \rangle  = \delta (t -t') dt $.
In all that follows we set $\mu =0$ for simplicity.
$P$ is the probability density of $\Omega$ conditioned on
$\Omega_0$ (which can be chosen to be $0$). Following the lines of standard stochastic calculus,
$P$ evolves according to a nonlinear Fokker-Planck equation of order $2-q$,
namely
\begin{equation}
\frac{\partial P }{\partial t} = \frac{1}{2} \frac{\partial^2}{\partial \Omega^2} P ^{2-q}
\end{equation}
Exact time-dependent solutions of this equation can be found \cite{Tsallis&Bukman} and are of the form of Tsallis
 distributions
of entropic index $q$ (equivalent to Student distributions)
\begin{equation}
\label{eq:Ptsallis}
P(\Omega_t \mid \Omega_0 ) = (a_q t^{\gamma} +  \frac{b_q}{t}(\Omega_t-\Omega_0 )^2)^{\frac{1}{1-q}}
\end{equation}
with
\begin{eqnarray}
\label{eq:aq}
a_q & = & ((2-q)(3-q)c_q)^{\gamma}\\
b_q & = & \frac{q-1}{(2-q)(3-q)} \label{eq:bq}
\end{eqnarray}
and
$\gamma = {\frac{q-1}{3-q}}$.
The $q$-dependent constant $c_q$ is given by $
c_q = {\frac{\pi}{q-1}} \frac{\Gamma^2(\frac{1}{q-1}-
\frac{1}{2})}{\Gamma^2(\frac{1}{q-1})}$.
From equation Eq (\ref{eq:y}) with $\mu=0$ it is evident that 
\begin{equation}
\Omega_t = \frac{y_t }{\sigma}
\end{equation}
so in the following we refer to $y$ as our variable of interest.

The index $q$  corresponds to the entropic Tsallis index within the general thermo statistics framework \cite{Tsallisentropy}.
It is also related to the degrees of freedom in a Student distribution.
If $q =1$, we recover a standard log-normal process for stock returns. If $q > 1$, the effective volatility
of the process is controlled by $P$ raised to a negative power. Therefore, fluctuations are enhanced if extreme
values of $y$ are realized, and they are more moderate if $y$ is less extreme {\it relative to the information 
available at time $0$.} This gives rise to
fat tails and bursts in the effective volatility of the process. 
Indeed, for $q > 1$ $P$ of Eq(\ref{eq:Ptsallis}) exhibits heavy tails. In particular, $q =1.5$
well-models the distribution of empirically observed log stock returns, reproducing the
observed cubic decay of the cumulative distribution of returns over short time scales.

The model here implies that the ensemble distribution of $y_t-y_0$  is of the Tsallis form
 for all times. It is harder to say something  about the distribution of increments over time $l$, $y_{t+l} - y_t$.
However, if we plot out  histograms of these increments we see that they have statistical properties consistent with 
empirical observations of log stock returns. But because the process explicitly depends on the initial 
value $y_0$ at 
time $0$, and therefore also explicitly on the time $t$, it is clearly not an optimal  model of real returns,
although it captures several realistic features.

Written in a slightly different notation so that the time $t_i$ is denoted by  $i$ and the time at a discrete 
increment $\Delta t $ later is denoted by $t_i + \Delta t = t_{i+1}$,
\begin{equation}
y_{i+i}= y_i +  \bar{\sigma}_i \Delta \omega_i
\end{equation}
where the volatility $\bar{\sigma}$ is defined as 
\begin{eqnarray}
\bar{\sigma}^2_i &=& 
\sigma^2 P(\Omega_i \mid \Omega_0)^{1-q} \\
&=&  (a_{i,0} + b_{i,0} (y_i - y_0)^2)
\end{eqnarray}
where we have replaced $ y_i = \Omega_i/\sigma$ 
and define
\begin{eqnarray}
a_{i,0}&  = &  a_q\sigma^2 i^{\gamma} \\
b_{i,0} & = & b_q i^{-1}
\end{eqnarray}
We denote
\begin{equation}
\label{eq:pij}
P_q(i,0) = (a_{i,0}  + b_{i,0} (y_i - y_0)^2)^{\frac{1}{1-q}}
\end{equation}
But for a multiplicative pre-factor, this is the conditional probability density 
of observing $y_i$ at time $i$ given $y_0$ at time $0$.
In the notation above it is clear that the process with volatility $\bar{\sigma}$ is simply a time- and state-dependent standard stochastic
equation. It just so happens that the solution yields a probability distribution for the variable $y$ that
is of the Student or Tsallis form.

\subsection{The multi-time scale  model}

As a more realistic approach, we propose the following model, a simple generalization of
the  one we previously proposed.  We  write the  volatility as
\begin{equation}
\bar{\sigma}_i^2 =  \frac{1}{N} \sum_{j=i-N}^{i-1} \bar{\sigma}_{i,j}^2
\end{equation}
where $N$ is the size of some past moving  window which contributes
to the effective volatility of the price, and
\begin{eqnarray}
\label{eq:barsigma}
\bar{\sigma}_{i,j}^2  
&=& P_q(i,j)^{1-q}\\
&= &(a_{ij} + b_{ij} (y_i - y_j)^2)
\end{eqnarray}
This leads to 
\begin{equation}
\label{eq:mt}
y_{i+i}  = y_i +  \frac{1}{\sqrt{N}} ( \sum_{j=i-N}^{i-1} \bar{\sigma}_{i,j}^2)^{\frac{1}{2}} \Delta \omega_i
\end{equation}
Here, $P_q(i,j)$  is of the form Eq (\ref{eq:pij}), with the argument $i,0$  replaced by $i,j$ 
and where 
\begin{eqnarray}
\label{eq:aij}
a_{i,j}&  = &  a_q \sigma^2 (i-j)^{\gamma} \\
\label{eq:bij}
b_{i,j} & = & b_q (i-j)^{-1}
\end{eqnarray}

Since market participants  are reacting to information on many different time horizons, the effective volatility of 
the process for $y$ is modeled here as the sum of contributions from all the different time horizons.
 In other words, from each time scale $i-j$,
there is a collective feedback into the system based on how extreme 
the movement of $y_i$ is perceived on the relevant time scale. 
If each  of the feedback processes on the individual time scales $i-j$ were independent, in the sense that only the
dynamics across that time scale would feed back into the system, then
$P_q$  would be  the probability density of the variable 
$y_i$ conditioned on an initial value $y_j$ at
 time $j$.
(Note that one possible more general formulation of this model would be to replace $P_q(i,j)$ with the true probability
density $P(y_i \mid y_j) $ of the model. But it is non-trivial to solve for this quantity  based on the 
 non-Markovian situation at hand, and so we choose instead to study the current simplified scenario). 
Thus, across each time horizon, the contribution to the  volatility follows the same dynamics
as an independent  statistical feedback process on that time scale. Our initial model of 
\cite{prl_borland,qf_borland} is recovered if only one time horizon is assumed relevant.

In the current form, the  memory in the system is described by an equal weighted moving average  that does not decay. We can easily modify this,
by introducing other types of averaging kernels $w_{ij}$ resulting in 
\begin{equation}
\label{eq:wweights}
dy_i = \frac{(\sum_{j= i-N} ^{i-1} w_{ij} (a_{ij} + b_{ij}(y_i-y_j)^2))^{\frac{1}{2}}}{\sqrt{\sum_{j=i-N}^{i-1} w_{ij} } } d \omega_i
\end{equation}
One possible choice of weights which we study in this paper is 
exponential decay $w_{ij} = \exp(-\lambda (i-j))$. If we let $N \rightarrow \infty$ this model represents pure exponential decay. For finite $N$ it describes a model with a cut-off exponential decay.  

Note that in this process there is only one driving noise acting at time $i$, namely the Gaussian random variable
 $\omega_i$. Via the statistical feedback process on different time-horizons,
the effective volatility at time $i$ is composed as the sum of a series of apparently random variables each 
the result of a statistical feedback process on 
the corresponding time scale.  Since we know the dynamics of each of these variables, the hope is that we can
perform useful calculations related to the statistics of $y$.

\section{Theoretical Calculations}

\subsection{The volatility}

We shall see that it is straightforward to calculate the second moment, or volatility, of this process.
We insert  $\bar{\sigma}$ as in  Eq(\ref{eq:barsigma}) into Eq (\ref{eq:mt}) and take the continuous time 
limit
 where 
\begin{eqnarray}
\label{eq:cont}
i & = &t_i \rightarrow t \\
 j& = &t_j \rightarrow s \nonumber \\
N & = & t_N \rightarrow T \nonumber 
\end{eqnarray}
 to get 
\begin{equation}
\label{eq:dyint}
d y_ t= \frac{1}{\sqrt{T}}( \int_{t-T}^{t }  
 (a_{t,s} + b_{t,s} (y_t - y_s)^2)ds )^{\frac{1}{2}}   d\omega_i
\end{equation}
Integrating over all past times $s$ which contribute to the process yields an effective volatility. 
We see that it must contain some sort of memory or volatility correlation.
Explicitly, for the short time volatility correlation,
\begin{eqnarray}
\langle   dy_t dy_{t'} \rangle & =&\frac{1}{T}\langle    (\int_{t-T}^{t} (t-s)^{\gamma}a_q \sigma^2 +b_q (t-s)
^{-1}(y_t -y_s)^2) ds ) ^{\frac{1}{2}}\nonumber \\ & & (\int_{t'-T}^{t'} 
((t'-t_s')^{\gamma}
a_q \sigma^2 +b_q (t'-s')
^{-1}(y_{t'} -y_{s'})^2) dt' ) ^{\frac{1}{2}} d\omega_t d\omega_{t'}\rangle \nonumber\\
&=&\frac{1}{T}\langle    \left( \int_{t-T}^{i}( (t-s)^{\gamma}a_q \sigma^2 +b_q (t-s)
^{-1}(y_t -y_s)^2)ds \right) \rangle  dt\nonumber \label{eq:dyt} \\
\end{eqnarray}
where we have used the property that $ \langle   d\omega_t d\omega_{t'}\rangle  =  \delta(t-t')dt$.

We evaluate this integral self-consistently by replacing 
 $(y_t - y_s)^2$ with its expectation. We obtain
\begin{equation}
\label{eq:yt2}
\langle    dy_t^2\rangle  =  \frac{a_q \sigma^2T^{\gamma}}{(\gamma+1)(1-b_q)} dt =  A_T dt 
\end{equation}
which does not depend on  absolute time. The fact that squared increments $\langle    dy_t^2\rangle $ are constants proportional to 
$dt$ implies a normally diffusive stationary process.  (Observe also that if one so wishes,  $\sigma^2 $ can always be chosen
 as $\tilde{\sigma}^2 T^{-\gamma}$ in order to cancel out the  $T$-dependent part of $\langle    dy_t^2\rangle $.)
Our result is exact, and valid as long as $b_q <    1$.

Very similar results are obtained  if we have an exponential decay
rather than a sharp cut-off, as in Eq (\ref{eq:yywweights}).
In the continuous time limit keeping the
notation of Eq(\ref{eq:cl}) one obtains
\begin{equation}
\langle    dy_t^2\rangle  = \frac{
\langle     \int_{t-T}^t \exp(-\lambda(t-s) (\sigma^2 a_q t^{\gamma} + b_q t^{-1}(y_t-y_s)^2) ds\rangle  dt}{\int_{t-T}^t \exp(-\lambda(t-s)) ds }
\end{equation}
Replacing $(y_t-y_s)^2$ inside the integral by its expectation, we solve self-consistently to obtain
\begin{equation}
\langle    dy_t^2\rangle   = \frac{\sigma^2 a_q \Gamma(\gamma+1)}{(1-b_q) \lambda^{\gamma}}dt = A_{\lambda} dt
\end{equation}
 where $\Gamma$ is the standard Gamma function and $T \rightarrow \infty$.

\subsection{Volatility-volatility correlations}

We can also look at the correlation of squared price differences  of this process, which in discrete notation can
 be written as
\begin{equation}
\label{eq:dydy}
\langle    \Delta y_i^2 \Delta y_{i+l}^2\rangle  - \langle    \Delta y_i^2\rangle \langle    \Delta y_{i+l}^2\rangle    = 3\Delta t^2( \langle     \bar{\sigma}_{i}^2 \bar{\sigma}_{i+l}^2\rangle 
 - \langle    \bar{\sigma}_i^2\rangle \langle    \bar{\sigma}_{i+l}^2\rangle )
\end{equation} 
If $l=0$ the two are perfectly correlated. If $l = N $ the two are perfectly uncorrelated  
(if we restrict our analysis to the accuracy  of the leading order in $q-1$, which means that we
 ignore the fact that $y_{i+N}$ itself depends on past values of $y$ in proportion to a $q-1$ factor).
 This is assumed to be the case in the discussion that follows (see also \cite{comment}). In that setting,
the correlation decays from the one extreme to the other as the lag $ l$ increases from $0$ to $N$.
Correlation is due to the fact that  there is a common set of histories which both $\Delta y_i^2$ and
 $\Delta y_{i+l}^2$ depend
on. The quantity we are interested in is the autocorrelation of volatilities,
namely  $C_{i,i+l} = \langle    \bar{\sigma}_i^2 \bar{\sigma}_{i+l}^2\rangle  - \langle    \bar{\sigma}_i^2\rangle \langle    \bar{\sigma}_{i+l}^2\rangle $, 
as seen here:
\begin{eqnarray}
C_{i,i+l} & =  & \langle     ( \sum_{j=i+l-N}^{i-1} \bar{\sigma}_{i,j} + \sum_{j= i-N}^{i-N+l} \bar{\sigma}_{i,j}  )
(\sum_{j= i + l -N}^{i-1} \bar{\sigma}_{i+l,j}  + \sum_{j=i}^{i+l}  \bar{\sigma}_{i+l,j})\ -\langle \bar{\sigma}_i^2\rangle \langle \bar{\sigma}_{i+l}^2 \rangle
  \nonumber \\
& = &\langle    ( C  + D)(\tilde{C} + \tilde{D})\rangle   - \langle \bar{\sigma}_i^2\rangle \langle \bar{\sigma}_{i+l}^2 \rangle \label{eq:C}
\end{eqnarray} 
where the notation should be obvious.
 $D$ and $\tilde{D}$ correspond to the  contributions  of the volatility-volatility correlation which are uncorrelated
with each other and with both $C$ terms. The $C$ terms on the other hand are correlated with each other.

In Eq(\ref{eq:C}),  the terms $\bar{\sigma}$ approach constants as  $q \rightarrow 1$, so the dependency on the random 
variables $y_i - y_j$
disappears and the problem becomes trivial.
For $q>  1$, as $l$ increases the common set of past values of $y_j$ which the two
 volatilities depend on decreases. This leads to a decrease in the volatility-volatility correlation at the rate $U(l)$.
In our case, when $l=0$ there is full correlation and this quantity is at a maximum. When  $l >   N$, there is complete
  de-correlation and this quantity should be zero.

From Eq(\ref{eq:C}) follows that the uncorrelated contribution  to $\langle    \bar{\sigma}_i^2 \bar{\sigma}_{i+l}^2\rangle $
can be written as
$\langle      \tilde{D} \bar{\sigma}_i^2 + D \tilde{C} \rangle  $. This is the amount that gets subtracted away from the  $\langle     \bar{\sigma_i}\rangle  
\langle     \bar{\sigma}_{i+l}\rangle $
term of the $C_{i,i+l}$ equation. Therefore, the rate at which the volatility autocorrelation decreases can be expressed as
\begin{equation}
\label{eq:ul}
 U(l) = \langle    \bar{\sigma}_i^2 \rangle \langle    \bar{\sigma}_{i+l}^2\rangle  - \langle      \tilde{D} \bar{\sigma}_i^2 + D \tilde{C} \rangle  
\end{equation}
This expression can be evaluated explicitly both for constant and exponentially decaying weights, as presented in the Appendix. 
In Figure 4 and the following paragraphs, the results of this calculation, namely  Eq (\ref{eq:cl}) and Eq (\ref{eq:yywweights}) together with Eq (\ref{eq:Dexp}) – Eq (\ref{eq:incAlambda})  are discussed in
more detail.

\section{Simulations and Numerical Results}

We created numerical simulations of the process (always with $\sigma = 1$ for simplicity), an example of which is shown in Figure 1. 
From this plot, non-Gaussian features such as bursts of volatility are apparent.
Not only do we see this volatility clustering by eye but also in the plot of Figure 2  where the 
variogram of volatility is depicted, defined as $v = \langle     \left(\ dy_t^2 - dy_{t+l}^2 \right)^2\rangle  $.
 For a random walk with no memory in the volatility, one would 
obtain a straight line. On the other hand, there is significant correlation  in the volatility  for $q=1.5$ and an
exponential decay.  Qualitatively similar results are obtained for other $q > 1$, 
and other cut-off lengths, with  constant or exponentially decaying weights.

\begin{figure}[t]
\psfig{file=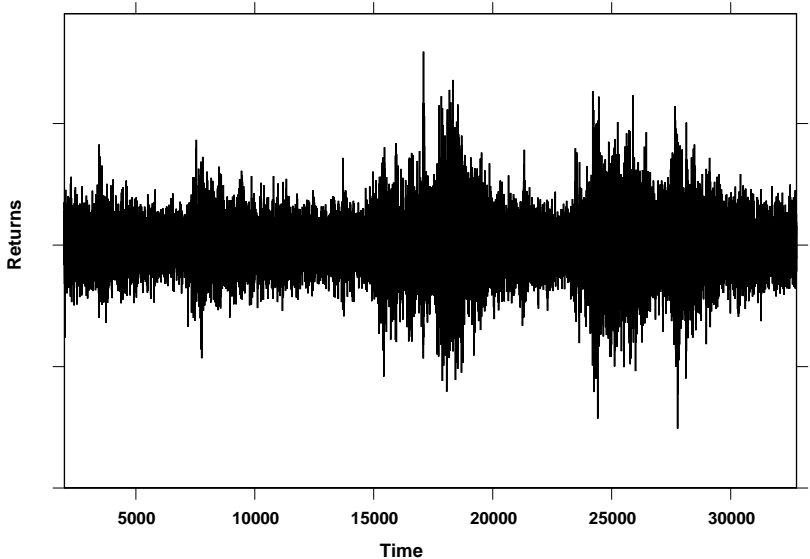,width=4.5 in}
\caption{\footnotesize
A typical path of the multi time-scale non-Gaussian model with $q=1.5$, equal weighted window with cut-off at $N=2000$ 
(which corresponds to $T =  140 $ days using $dt = 1./14$  as the simulation step)
is shown. Very similar results are obtained for a wide range of cut-off lengths, and choice of exponential weight.
}
\end{figure}

\begin{figure}[t]
\psfig{file=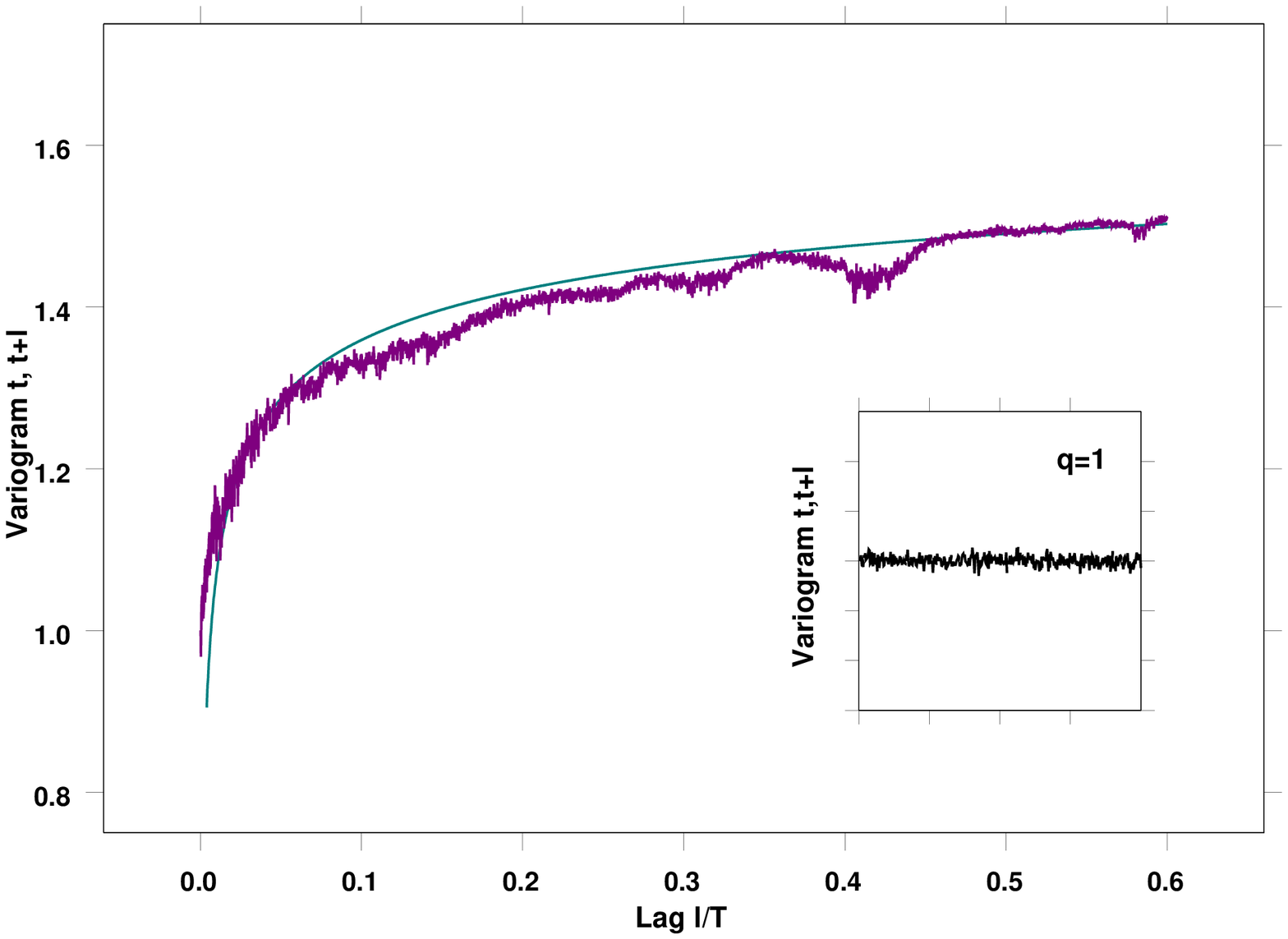,width=4.5 in}
\caption{\footnotesize
A volatility variogram shows the correlation of volatility across time scales.
Here for $q= 1.5$ and an exponentially decaying  weight with half-time at 
$l = 500 dt \approx 36 $ days  and $T = 5000 dt \approx  1 $ year.
The solid line corresponds to a theoretical curve assuming the variogram goes as Eq(\ref{eq:vgram}) with 
$g= l^{-0.22}$ which is a proxy for  the behaviour of  real stock returns.
The inset depicts the $q=1$  case (standard Black-Scholes log-normal model), that shows no correlation or memory.}
\end{figure}

Another feature of stock returns is that the  kurtosis approaches zero slowly as the time lag of the returns
increase. Over short time-scales, returns 
have a high kurtosis (are very non-Gaussian with fat tails) but as the time scale increases, the kurtosis
tends to vanish and returns appear more Gaussian.  
 The rate of this decay of kurtosis can roughly be approximated by $U(l)$, the rate of decay of $C_{i,i+l}$ \cite{bouchaudpotters}. 
In Figure 3 we plot the
empirical kurtosis 
calculated from simulations of the process using an equal weighted window, together with a plot of $U$ of Eq(\ref{eq:cl}).
 Also shown is the
line depicting $l^{-.22}$ which we use as a proxy for empirical decay of the kurtosis of real stock returns \cite{bouchaudpotters}. 
It is clear that our model indeed has a very slowly decaying kurtosis. In fact, with an equal weighted window  it decays even slower than what is observed for 
real stock returns (for lags sufficiently smaller than the cut-off time).
However, if we include an exponential weight rather than a constant one, then 
    the decay of the kurtosis is more realistic.

\begin{figure}[t]
\psfig{file=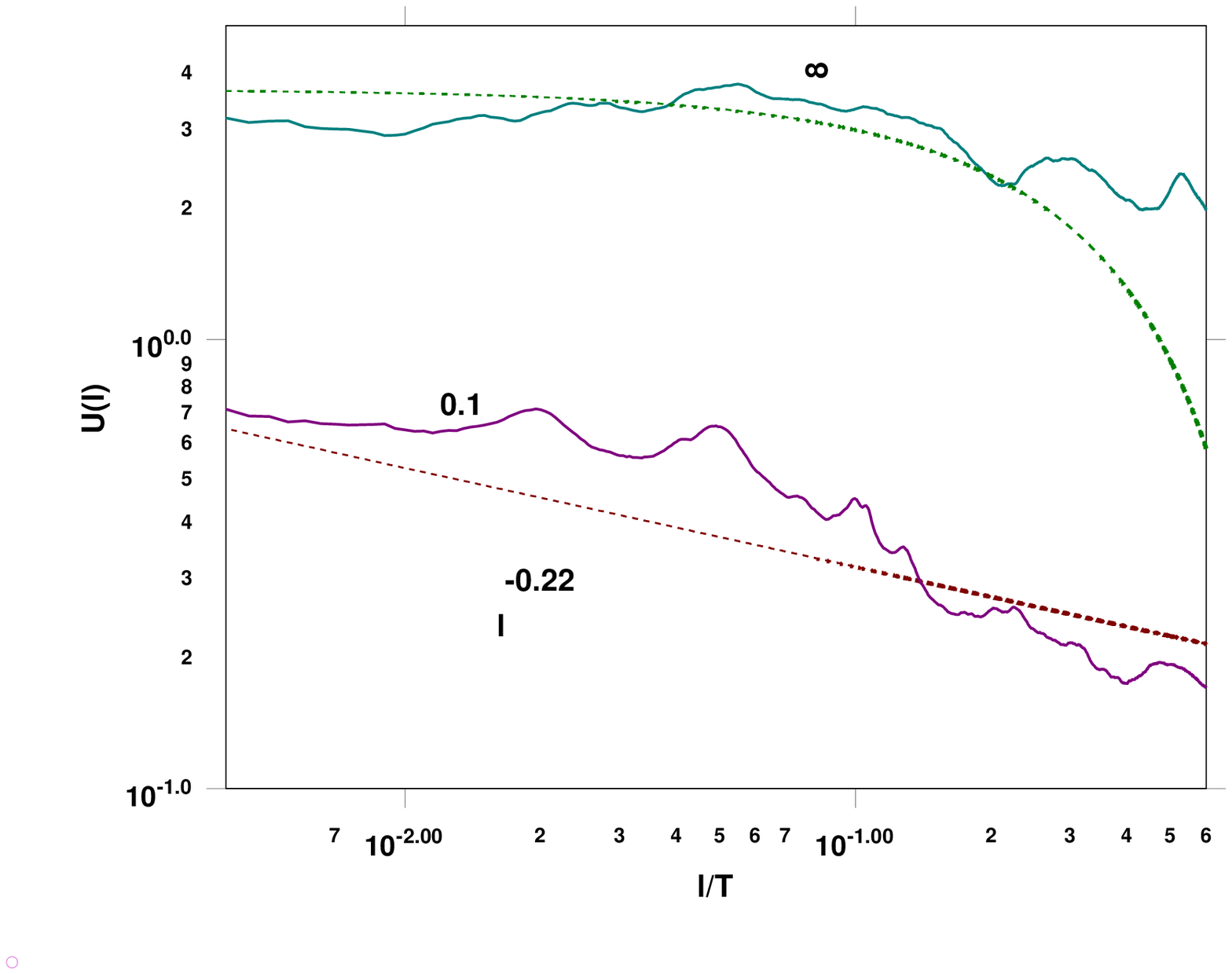,width=4.5 in}
\caption{\footnotesize
The average kurtosis calculated from (top curve) a simulated time series of the process with constant equal weights,
together with the  $U(l)$ Eq (\ref{eq:cl}). The kurtosis decays slower that that of real stock returns which
goes as $l^{-0.22}$,also shown.
By using exponential weights (here of  half-time corresponding to $h = l_{0.5}/T =  0.1$), simulations
yield a kurtosis that matches well to real stocks.
 }
\end{figure}

In fact, it  is easy to see that it is the rate of decay of the 
memory across time-scales which controls the rate of decay of the   kurtosis.
This is explored in Figure 4.  The parameter which we vary is the half-time of the exponential weight defined as $l_{0.5} = log(2)/\lambda$.  For  numerical reasons, we kept $N$ large yet finite in all simulations, at a value of $N = 5000$, corresponding to $T \approx 1$
year  with our choice of $dt$. Consequently, it is convenient to rescale $l$ by $T$, and study the behaviour of the process as $h = l_{0.5}/T $ varies from $ \infty$ (a constant equal weighted scenario) to $0$ (no memory at all).
 The rate of decay of the kurtosis varies from that of 
Eq(\ref{eq:cl}) which 
is close to constant for times less than the cut-off time $T$, to roughly  $l^{-1}$ which is the result we expect for perfectly uncorrelated
Gaussian random variables. From the plot in Figure 4b it seems that choosing a value of $h$ in the range $0.02 = 0.1$ reproduces
a decay of kurtosis reasonable when compared to that of real stock returns, if we look at the region  $l \ll T$. These values correspond to half-times of 
7 days and 36 days,  respectively, with the choice of simulation parameters used.

\begin{figure}[t]
\psfig{file=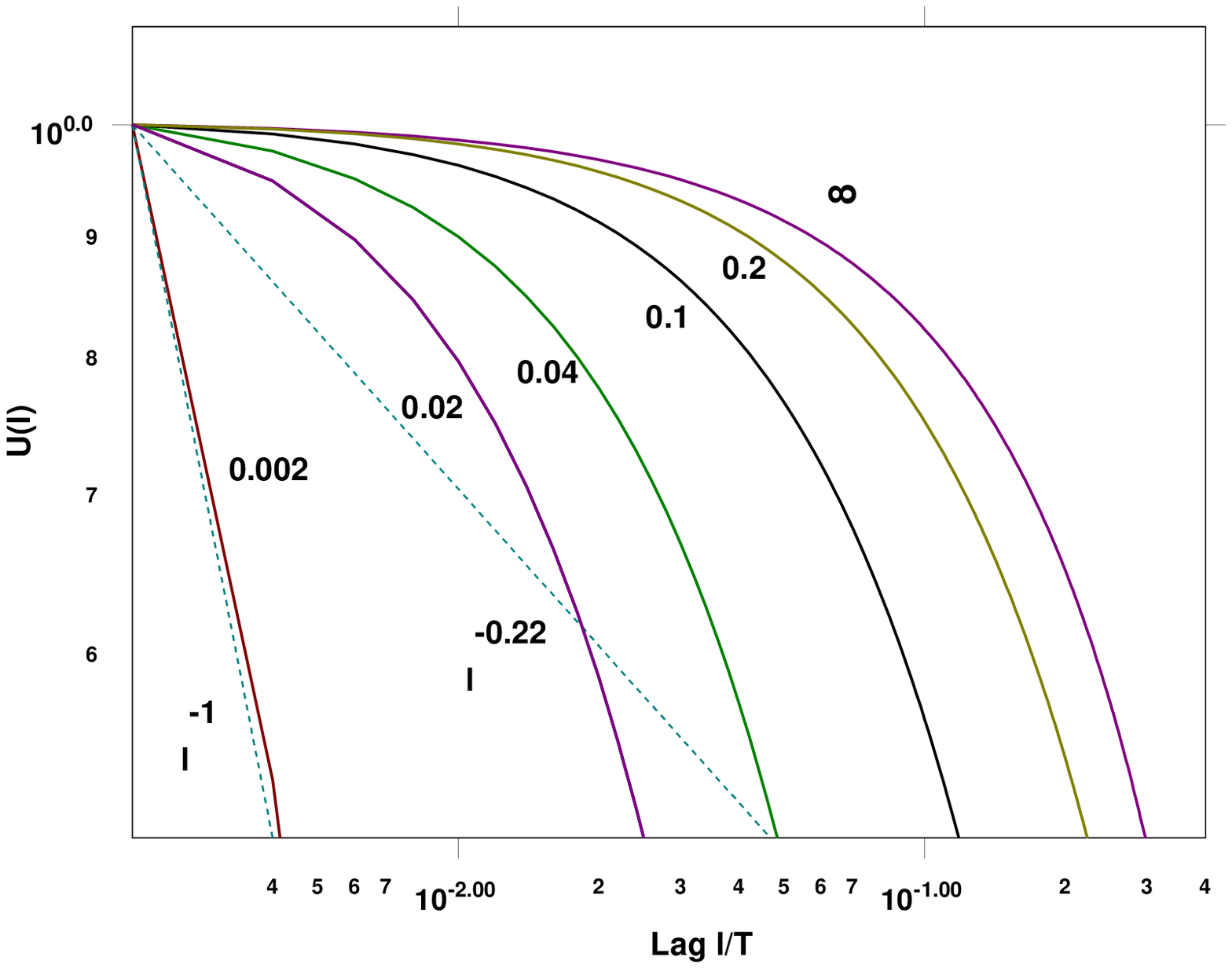,width=2.5 in}
\psfig{file = 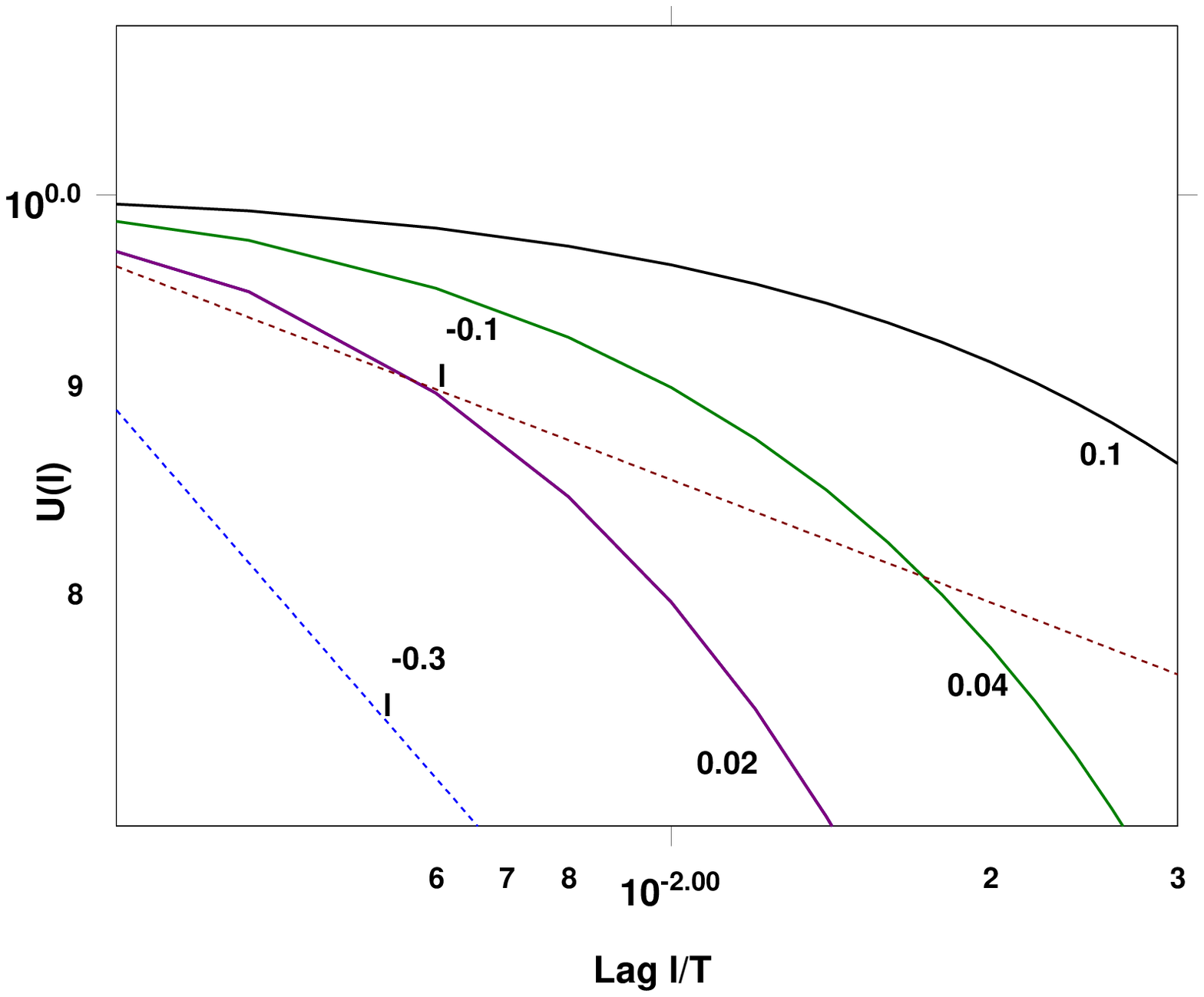, width=2.5 in}
\caption{\footnotesize
a) The average kurtosis decreases slowly across time scales following Eq(\ref{eq:cl}). Here we keep $T$ fixed 
 and vary the relative 
half-time $h = l_{0.5}/T$
from $h = \infty $ (equal weights with no decay) toward $h = 0$. We also show  the line of slope $-0.22$
(a proxy for behaviour of real stocks) and 
a line of slope $-1$ (a proxy for uncorrelated Gaussian variables) for comparison . (The fast decay as $l/T $
 increases is due to the cut-off, and is
reduced if $T\rightarrow \infty$.) All curves are normalized to start at $1$.
b) This plot is a sub-set of the one above, together with two lines of slope ${-0.1}$ and ${-0.3}$. These lines 
quite well encompass the range of slopes of the curves for $h = 0.02 - 0.1$, shown also. On average, the slope is
close to that observed for real stocks ($\approx -0.2$). 
}
\end{figure}

Note that in our model, the  behaviour of $U(l)$   depends  only trivially on the value of the parameter $q$ (which essentially controls the
 strength of the feedback into the system)
   via the additive terms due to $a_{ij}$  and the 
pre-factor $b_q$ which multiplies the  $y_i-y_j$ terms. On a log-log plot, the slope of $U(l)$ is constant for all $q$,
the magnitude however vanishing as $q  \rightarrow 1$. If we normalize $U(l)= 1$
for $l=0$  for different values of $q$, then indeed all the $U(l)$ curves collapse onto each other.
 Thus, it is not $q$ but rather the  temporal behaviour  of the  weights $w_{ij}$  that controls the rate of decay of $U$. In other 
words, the slow decay of volatility-volatility correlations is  a signature of the memory of the system.

Let us return our focus to the plot in Figure 2.
We can compare the form of this volatility-volatility variogram with known results of real stock markets 
\cite{bouchaudpotters}. In particular, Bouchaud and Potters show that 
\begin{equation}
\label{eq:vgram}
v = \langle    \left(\ dy_t^2 - dy_{t+l}^2 \right)^2\rangle  \propto v_1 - U(l) v_2
\end{equation}
where $v_1$ and $v_2$ are case-specific constants.
We already mentioned that  they find that $U(l) \propto l^{-.22}$ fits real stock returns quite well. This variogram was obtained
from a simulation with $h = 0.1$ which from the discussion in the above paragraph should reproduce a somewhat realistic decay
of kurtosis, and therefore also a realistic variogram. Indeed, the solid line in Figure 2 is obtained from  a fit of 
Eq(\ref{eq:vgram}) with coefficients $v_1 = 1.8$ and $v_2 = 1.73$. The agreement is quite close.

Figure 5 serves to illustrate yet again the feature of decreasing kurtosis. We have plotted out histograms of returns from 
a simulation of
the process with $q=1.5$ and an exponential half-time corresponding to $h = 0.1$.  The distribution is tent-like for 
small time lags.
As the time lag increases, the distributions become more Gaussian \cite{bouchaudpotters, stanley,osorioetal}.
A systematic analysis of these distributions over time will be part of future work \cite{jpandme}.
Finally, we also plot out the distribution of the instantaneous volatility calculated from the same simulation.
 As a proxy for the volatility we have used squared returns averaged over 3 time increments.
In Figure 6 we  show the empirical distribution of the logarithm  of the volatility from such a calculation.
It is very close to a  Gaussian distribution, as is quite apparent from the parabolic shape in this semi-log 
representation. This is encouraging, because it is well-known that real stock return volatility is well modeled by a
log-normal distribution.

\begin{figure}[t]
\psfig{file=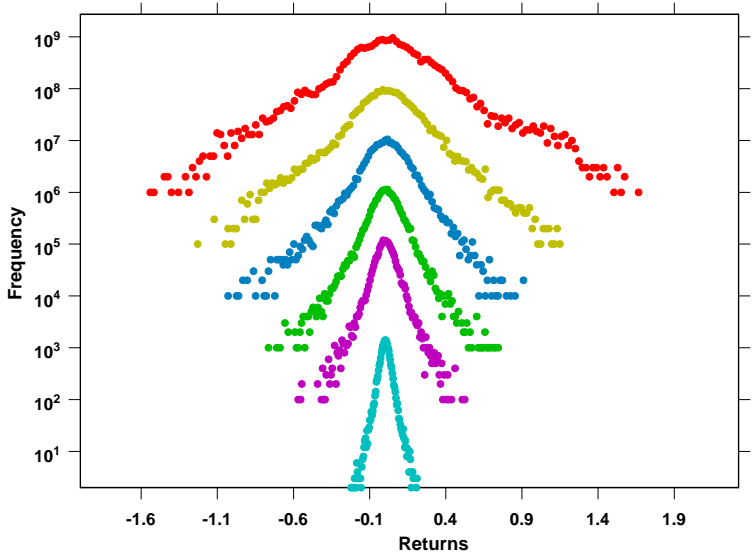,width=4.5 in}
\caption{\footnotesize  Shown here are histograms of returns $y_{t+l} - y_t$ for different time lags $l$.
From bottom to top corresponds the lags are  $l=1,2,4,8,16$ and $32$ multiples of the simulation time-step. 
For small $l$ the distribution is highly non-Gaussian,  becoming  more Gaussian as the time scale increases.
This is  consistent with a decreasing kurtosis.}
 \end{figure}

\begin{figure}[t]
\psfig{file=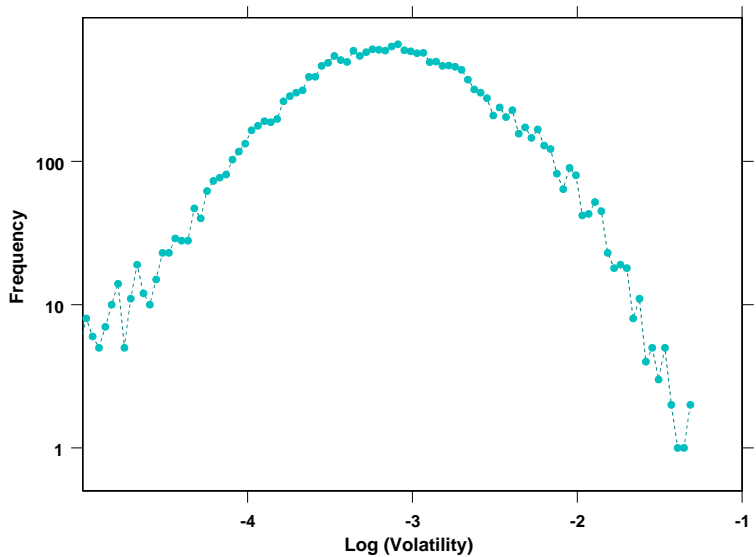,width=4.5 in}
\caption{\footnotesize
The distribution of the effective volatility appears 
similar to that of real stock returns, very close to a log-normal distribution, implying
that the log-volatility follows a normal distribution.
This is evident from the parabolic shape of the frequency plot shown here in semi-log representation.
}
\end{figure}

It is worth mentioning  that in  principle one might be able to relax the assumptions of our initial model and 
still obtain 
processes with very similar features.
Instead of arguing that at each time horizon $i-j$, the feedback process depends  on $P_q(y_i \mid y_j)$,
we could argue that it depends on $P_q(y_i \mid \bar{y}_j)$ where $\bar{y}_j$ can be any appropriate 
reference price, average of past prices, mean of the distribution of prices and so on.  Exploring this in more detail
is however beyond the scope of this paper.

\section{Conclusions}

In conclusion we have proposed a  stochastic process with just one source of Brownian noise,
 yet the instantaneous volatility is the result of a kind of statistical 
feedback across many time scales. This deterministic-volatility model  captures
many features observed in time series of real stock returns. These features or {\it stylized facts} include
a log-normal distribution of volatility, together with volatility clustering  and a  kurtosis which decreases 
slowly over time. We find that the rate of 
this decay can easily be matched  to that observed empirically for stock returns, and depends on the memory inherent
in the system.

Our model is parameterized by the index $q$, which corresponds to the  degree of feedback into the system.
For $q=1$, there is no feedback and the process is identical to a standard Brownian motion. For $q > 1$ there
is a feedback, and there  is a contribution to the instantaneous volatility  from multiple time horizons, such that 
the fluctuations depend on the probability of the latest observation at time $i$ with respect to past observations 
on those different
time horizons $j$. We model the probability densities  as Tsallis (or Student) distributions, which  they would be if each
 feedback process were
independent. 
 Mathematically, this non-Markovian process contains a sum over return differences weighted by a function  
of the  
time difference $i-j$.  The exact form of this weighting controls the strength of the memory in the system.
In this paper we explored the effects of both constant and exponentially decaying weights. For these two 
scenarios we  could calculate the
second moment as well as $U(l)$, the decay of the fourth moment correlation. 
 We found that on a log-log plot,
the slope of this function $U$ is constant for all q but varies according to the strength of the memory in the system.  
 The function contains a multiplicative  $q$-dependent pre-factor which  goes to zero as $q$ approaches $1$, resulting in
 a vanishing kurtosis
for that case. Other functional forms of the weighting kernel, such as a power-law or a weight which emphasizes 
natural time scales such as days, weeks or months, will be explored in future work \cite{jpandme}.

These theoretical results in addition to  the analysis of simulations  showed that 
this model well-captures many stylized facts of real financial data. Furthermore it is simple and rather intuitive.
The premise (that the volatility is a deterministic function of returns over different time horizons) is consistent with
recent empirical analysis \cite{zumbachlynch,zumbach}.
In a forthcoming paper \cite{jpandme} , we shall analyze this  and a related model in further detail, both theoretically 
and with respect to empirical evidence of the underlying assumptions and possible  predictions. A great challenge is to try
and solve explicitly the distribution of the variables $y_{t+l} - y_t$.

As a final remark we reiterate  that in the case where there is only one relevant time, the current model reduces 
to the one 
which we previously proposed \cite{qf_borland}. That model has been used for developing a theory of option pricing and
another focus of future work will be to try and extend that theory to incorporate the multiple time horizons
as introduced in the present framework. 

\appendix

\section{Appendix}
The quantity we would like to evaluate is Eq (\ref{eq:ul}),
namely
\begin{equation}
 U(l) = \langle   \bar{\sigma}_i^2 \rangle \langle   \bar{\sigma}_{i+l}^2\rangle  - \langle     \tilde{D} \bar{\sigma}_i^2 + D \tilde{C} \rangle  
\end{equation}

The $\bar{\sigma}^2$ terms
are both equal to the diffusion coefficient $A_T$ of Eq (\ref{eq:yt2}).
We now replace the $D$, $\tilde{D}$ and $\tilde{C}$  terms with the sums they correspond to. 
\begin{eqnarray}
\langle   D \rangle &= &\langle   \sum_{j= i-N}^{i+l-N } a_q (i-j)^{\gamma} + b_q (i -j)^{-1} (y_i-y_j)^2 \rangle \\
\langle   \tilde{D}\rangle  &=& \langle   \sum_{j=i}^{i+l}  a_q (i+l -j)^{\gamma} + b_q (i+l-j)^{-1}(y_{i+l} - y_j)^2\rangle  \\
\langle   \tilde{C}\rangle  &=& \langle   \sum_{j=i +l-N}^{i} a_q (i+l -j)^{\gamma} +  b_q (i+l-j)^{-1}(y_{i+l} - y_j)^2\rangle 
\end{eqnarray}
We take the continuous limit of these sums, using the notation of Eq (\ref{eq:cont})
To evaluate the integrals, we replace the $(y_i - y_j)^2$ terms by their expectations according to
Eq (\ref{eq:yt2}). We finally obtain
\begin{eqnarray}
\label{eq:cl}
U(l) &= &A_T^2 - A_T \left(\frac{a_q l^{\gamma+1}}{(\gamma+1)T} + b_q A_T \frac{l}{T} \right)  \\
&-& 
 \left( \frac{a_q}{(\gamma+1)T} (T^{\gamma+1} - (T-l)^{\gamma+1})  + b_q A_T \frac{l}{T} \right) \nonumber \\
& & \left(\frac{a_q}{ (\gamma+1)T} (T^{\gamma+1}-l^{\gamma+1}) + b_q A_T (T-l) \right) \nonumber
\end{eqnarray}

This result is valid for a constant equal weighted window. However it is straightforward to generalize for the case 
of exponential weights. Defining $dy_i$ as in Eq (\ref{eq:wweights}) we obtain (after taking the continuous limits 
and integrating):
\begin{eqnarray}
\langle   D\rangle  &=& b_q A_{\lambda}(\exp(-\lambda (T-l) - \exp(-\lambda T)) \nonumber \\
&  +&  \frac{a_q}{\lambda^{\gamma}}( \Gamma(\gamma+1,\lambda(T-l))
-\Gamma(\gamma+1,\lambda l)) \label{eq:Dexp}\\
\langle   \tilde{D}\rangle  &=& b_q A_{\lambda} (1-\exp(-\lambda l)) \nonumber \\&+& \frac{a_q}{\lambda^{\gamma}} ( \Gamma(\gamma+1) - \Gamma(\gamma+1,\lambda l))
\label{eq:Dtildeexp}\\
\langle   \tilde{C}\rangle  & = & b_q A_{\lambda} (\exp(-\lambda l) - \exp(-\lambda T))\nonumber \\
& +& \frac{a_q}{\lambda^{\gamma}} ( \Gamma(\gamma+1,
\lambda l) - \Gamma ( \gamma+1, \lambda (T-l)) \label{eq:Ctildeexp}
\end{eqnarray}
together with  
\begin{equation}
\label{eq:incAlambda}
A_{\lambda} = \frac{\sigma^2 a_q (\Gamma(\gamma+1)-\Gamma(\gamma+1,\lambda T) )}{(1-b_q) \lambda^{\gamma}}
\end{equation} 
 where $\Gamma(a)$ is the standard Gamma function and  $\Gamma(a,x)$ represents the  incomplete Gamma function. 
Here $\Gamma(a,x)$ corresponds to the incomplete Gamma function and $\Gamma(a)$ the standard Gamma function.
Inserting these equations into Eq(\ref{eq:ul}) yields
\begin{equation}
\label{eq:yywweights}
U(l) = A_{\lambda}^2 - \langle     \tilde{D}\rangle  A _{\lambda} - \langle   D \rangle  \langle   \tilde{C} \rangle  
\end{equation}
If we now  let $T$ go towards infinity in this calculation,  we obtain
 results for the case of pure exponential decay.

\section*{Acknowledgements}

The author thanks Jean-Philippe Bouchaud for many inspiring discussions, useful comments and suggestions. 
Benoit Pochart and Roberto Osorio have been very helpful in critical reading of this manuscript.
Jeremy Evnine is thanked for his continued support and useful comments. Interesting discussions with Gilles
Zumbach and J.-F. Muzy at the 2004 Leiden Workshop on "Volatility in Financial Markets" are also acknowledged.


\end{document}